 \documentclass[preprint2]{aastex}

\usepackage{epsfig}
\usepackage{epstopdf}

\usepackage{amssymb}

\shortauthors{F. Zagury}
\shorttitle{The interstellar extinction curve}

\begin{document}

\title{On the relationship between the continuum of interstellar extinction curves, the 2200~\AA\ bump, and the diffuse interstellar bands}

\author{Frederic Zagury}
\received{ }


\begin{abstract}
A previous article argued that the antagonism between sight-lines with and without a bump at 2200~\AA\ disappears, and that the observed properties of interstellar extinction can be globally understood, if it is accepted that scattered starlight contaminates the observed spectrum of reddened stars.
The present paper develops this new paradigm by providing a better understanding of the characteristics of this scattered light.
It further examines the consequences of this revision of interstellar extinction theory for interpreting  the 2200~\AA\ bump and the diffuse interstellar bands (DIBs).
Two implications are worth noting: (1) the effect of interstellar extinction on cosmic distance estimations needs to  be reconsidered; (2) it is unnecessary to invoke hypothetical particles, such as Poly Aromatic Hydrogenated molecules (PAHs),  to explain the peculiarities of interstellar extinction. 
Hydrogen, by far the most abundant constituent of interstellar clouds, should alone account for the three major features  of interstellar extinction: the departure of the continuum from linearity, the bump, and the DIBs.
\end{abstract}

\section{Introduction}
After decades of research and despite a now well-established set of observations, studies on interstellar extinction still lack a firm and internally-consistent theoretical framework.
The nature of the particles that would explain the  2200~\AA\ bump and the far-ultraviolet rise of extinction curves remains controversial.
In the visible wavelength range, the carrier(s) of the diffuse interstellar bands (DIBs) have still not been identified.

It is taken for granted in  interstellar extinction  studies that an observer sees the direct light from reddened stars  dimmed by different types of interstellar particles in the line of sight.
However, this assumption leads to contradictions between theory and observations at ultraviolet wavelengths (\cite{an13}, Sects.~\ref{ccm} to \ref{contra} this paper).

An alternative, the only one in fact,  would be to accept that a large amount of scattered starlight is mixed with the direct light we receive from reddened stars \citep{an13}.
This shift in perspective would unify existing observations in a consistent picture.
It would simplify the true extinction law of interstellar dust (which is linear over the whole spectrum, $\tau_\lambda \propto 1/\lambda^p$, $p\sim 1$), and explain the observed relationship between the extinctions in the optical and in the ultraviolet parts of the spectrum.
It would also explain why some extinction curves have no 2200~\AA\ bump and continue to be linear in the ultraviolet  spectrum (Fig.~\ref{fig:fig1}).
Finally, it would settle long-lasting controversies on the value of the ratio of total to selective extinction $R_V$ parameter (Sect.~\ref{cons}).

This alternative is itself not free from difficulties.
If light received  from a reddened star decomposes into  direct and  scattered light components, the latter can overwhelm the former.
At ultraviolet wavelengths a star would appear much brighter if it could be observed through an interstellar cloud of pure gas (no dust) than would be the case without the cloud in the line of sight.
The cloud (with no dust)  would amplify the signal from the background star and act more like a lens than as an extinction medium.
On observed extinction curves this effect is masked by the exponential extinction $e^{-\tau_\lambda}$ ($\tau_\lambda\propto 1/\lambda^p, \, p\sim 1$) that interstellar dust introduces.
It is recovered when the latter is removed (Fig~\ref{fig:fig2}).

Diffraction in the forward direction, the specificity of which is well known and recognized \citep{vdh, hecht}, should alone account for the strength of the scattering component in the spectrum of reddened stars.
However, no  theory  provides a straightforward fit to interstellar extinction curves.
Textbooks  \citep{hecht, stone} discuss and illustrate the pattern of forward diffracted light by randomly distributed particles, but their accounts remain at a qualitative level. 
The formula for the irradiance of the forward scattering by a slab of identical particles, derived by H.C. van de Hulst  in the 1950s, seems to anticipate the lensing effect of interstellar clouds, but it predicts ratios of the scattered to the direct starlight intensities that  are much too high, and also a different wavelength dependence than is actually observed (\cite{oc12} and Sect.~\ref{vdh}).

A particularity of forward scattering  in the case of interstellar extinction is the  impressive size,  a few thousand kilometers wide, that Fresnel zones can reach on astronomical scales.
The larger the Fresnel zone is at the position of the cloud, the greater the power available for scattering and the greater the proportion of scattered starlight in the spectrum of a reddened star.

It is the combination of complete forward scattering, large Fresnel zones, and identical particles (atomic and/or molecular hydrogen) that explains the very large proportion of scattered starlight in the spectrum of reddened stars and why observed extinction curves depart from linearity in the ultraviolet spectrum.
Thus restated, the issue raised by ultraviolet extinction curves highlights the necessity of shifting research on  interstellar extinction from an inquiry into the  chemistry of the interstellar medium to a problem of optics. 

This shift would reframe our understanding of the two major issues cited above, the 2200~\AA\ bump and the DIBs.
The bump, a singular figure in the ultraviolet part of extinction curves,  must be seen as an interruption of the scattered light component of the spectrum of reddened stars (Sect.~\ref{abs} and Fig.~\ref{fig:fig1}).

In 2005 the state of DIB research was summarized as follows:  \begin{quotation}In May of 1994, the first major conference focusing entirely on the baffling problem of identifying the carriers of the diffuse interstellar bands (DIBs) was held at the University of Colorado, Boulder. From even a perfunctory reading of the subsequently published book (Tielens \& Snow 1995) containing the invited papers presented at the meeting, one would gather that there was a growing sentiment among the conference participants that the DIB carriers would soon be shown to be large, complex molecules. 
More than ten years have now passed since the Boulder conference was held. 
Despite intensified, technically sophisticated efforts ... to identify such molecules made during this period by several world class teams of astronomers, chemical kineticists, and spectroscopists, the situation remains today exactly as it was in 1994 - \textit{i.e.} not a single band in the total DIB spectrum (now known to contain more than 250 bands) has been convincingly assigned to an optical transition of any molecule, atom, or ion!\footnote{Similar assessment are found in \cite{tielens08,draine09}.}   ... [Considering] that optical absorption spectroscopy is a fully developed field of science that has been extensively employed to probe atomic and molecular structure for a very long time, it is now perhaps reasonable to begin to suspect that the up-to-present total lack of success in identifying the carrier(s) of the DIB spectrum is the result of assuming that the bands in this spectrum are produced via the effect of \textit{linear absorption}.
 \citep{sorokin05}.\end{quotation}

As the sensitivity and the spectral resolution of observations increase, new DIBs (now over 500, see \cite{hobbs09}) continue to be found,  but, over ten years after the   \cite{sorokin05} paper and twenty years after the Boulder meeting,  the identification of the "large, complex molecules" that would explain the features  is  at a standstill.
Neither  fullerenes \citep{campbell15}, nor Poly Aromatic Hydrogenated molecules (PAHs), will account for the full DIB spectrum,  and the presence of these molecules in interstellar clouds raises  difficulties on account of their supposed abundance (Sect.~\ref{pah} and \cite{ubachs14}).
The situation is all the more perplexing because interstellar clouds sampled in extinction observations are known to be cold and near perfect vacuum regions of interstellar space, which leaves little room for a sophisticated chemistry.

Sorokin \& Glownia question whether a classical linear absorption can explain the origin of the DIBs. 
The two-photon absorption process they suggest (which implies that DIBs are created close to the stars) is no more convincing (Sect.~\ref{abs}), unless the continuum of scattered starlight in the spectrum of reddened stars provides the coherent source of light needed for two-photon absorption to occur.  
In this respect understanding the relationship among the ultraviolet continuum of extinction curves, the bump, and the DIBs becomes very important.

The first part of this article  (Sects.~\ref{pec} to \ref{oth}) deals with the continuum of extinction curves.
A special emphasis is given to a critical issue, the number of free parameters, or degrees of freedom, that interstellar extinction curves depend on.
The incompatibility between a theory that requires a minimum of six independent parameters (Sect.~\ref{fitz}), and observation, which shows that one is enough (Sect.~\ref{ccm}), does not seem to be recognized even in recent publications and textbooks (Sect.~\ref{ccmfitz}).
The remarkably  precise shape of extinction curves is also  difficult to accommodate with the presence in interstellar clouds of different types of particles, the proportions of which can fluctuate independently of one another \citep{greenberg00}.

The observed properties of the scattered-light component of extinction curves are analyzed from Sects.~\ref{contra} to \ref{cons}.
Sect.~\ref{oth} treats issues such as the limits of van de Hulst's formula and energy conservation.
Sect.~\ref{dib} emphasizes the link between the scattered-light component of extinction curves, the 2200~\AA\ bump, and the DIBs, and is continued in Sect.~\ref{abs} by a discussion on the  Sorokin \& Glownia mechanism to produce the DIBs.
 
The implications of these views on interstellar extinction theory for distance estimates and for ongoing research on hypothetical particles of the interstellar medium, such as the PAHs, are drawn in Sect.~\ref{mis}.

An appendix summarizes elementary principles and particularly useful results of diffraction theory.
The small vibrations of the hypothetical "particules \'eclairantes"  (or   "mol\'ecules \'ether\'ees" \cite[p.~202, 375]{fresnel}) that  Fresnel imagined to account for the propagation of light, which still inform modern presentations of classical diffraction theory \citep{feynman,born, hecht}, should find an even more realistic application in coherent Rayleigh scattering  than they do for the free propagation of light.
\section{Interstellar extinction curves} \label{ecdef}
Extinction curves $A_\lambda$ have been introduced as the magnitude of the ratio $F/F_{ref}$ of the spectrum $F$ of a reddened star to the spectrum $F_{ref}$ of a non-reddened star of the same spectral type: $A_\lambda=-2.5\log(F/F_{ref})+C$. 
Constant $C$  is eliminated by referring the extinction to a given wavelength \citep{rv}.
Extinction curves are usually normalized by the reddening $E(B-V)=A_B-A_V$ of the star.
The reddening $E(B-V)$ is a measure of the column density of interstellar matter between the star and the observer.
Normalized extinction curves measure  the extinction per unit reddening in the direction of the star.

In a pure extinction process $A_\lambda$  is nearly the wavelength-dependent optical depth $\tau_\lambda$ in the expression $F/F_0 =e^{-\tau_\lambda}$ of the attenuation of light traversing a medium.
It is a convenient parameter to characterize the particles along the line of sight and their column density.
If different particles or media are along the line of sight, $F/F_0$ is the product of the $e^{-\tau_\lambda}$ over the particles and media, and the extinction curve is the sum of the  extinctions $A_\lambda$.

These properties justify the longstanding choice of  $A_\lambda$ over $F/F_{ref}$ (given by observation) as the variable for interstellar extinction observations.
This  choice, however, impacts the representation of  extinction curves and  affects how they are interpreted.
The replacement of $F/F_{ref}$ by  $A_\lambda$ can be particularly misleading if the pure extinction of starlight is not the only process at work.

Both understandings ($\propto F/F_{ref}$ or $-2.5\log(F/F_{ref})+C$) can be used to designate an extinction curve since the context generally prevents confusion.
A 'linear' (or quasi-linear) extinction law means that the wavelength dependence of the extinction is $\tau_\lambda \propto 1/\lambda^p$ and equally that the attenuation of light is $F/F_{ref} \propto e^{-a/\lambda^p}$ ($p\sim1$).
Constant $a$ can be related simply to the reddening $E(B-V)$ for any value of the ratio of total to selective extinction  $R_V=A_V/E(B-V)=\tau_V/(\tau_B-\tau_V)$ \citep{rv2}.
\section{Observed properties of interstellar extinction curves} \label{pec}
Within $1$ to $0.4\,\mu$m in the visible wavelength range, all extinction curves follow a linear extinction law $\tau_\lambda \propto 1/\lambda^p$, $p\approx1$ ($F/F_{ref} \propto e^{-a/\lambda^p}$).
The linearity of the extinction tends to break under 400~nm, as if there was less extinction than predicted by the continuation of the linear law into the near-ultraviolet domain \citep{nandy64, rv}.

\begin{figure*}[]
\resizebox{1.85\columnwidth }{!}{\includegraphics{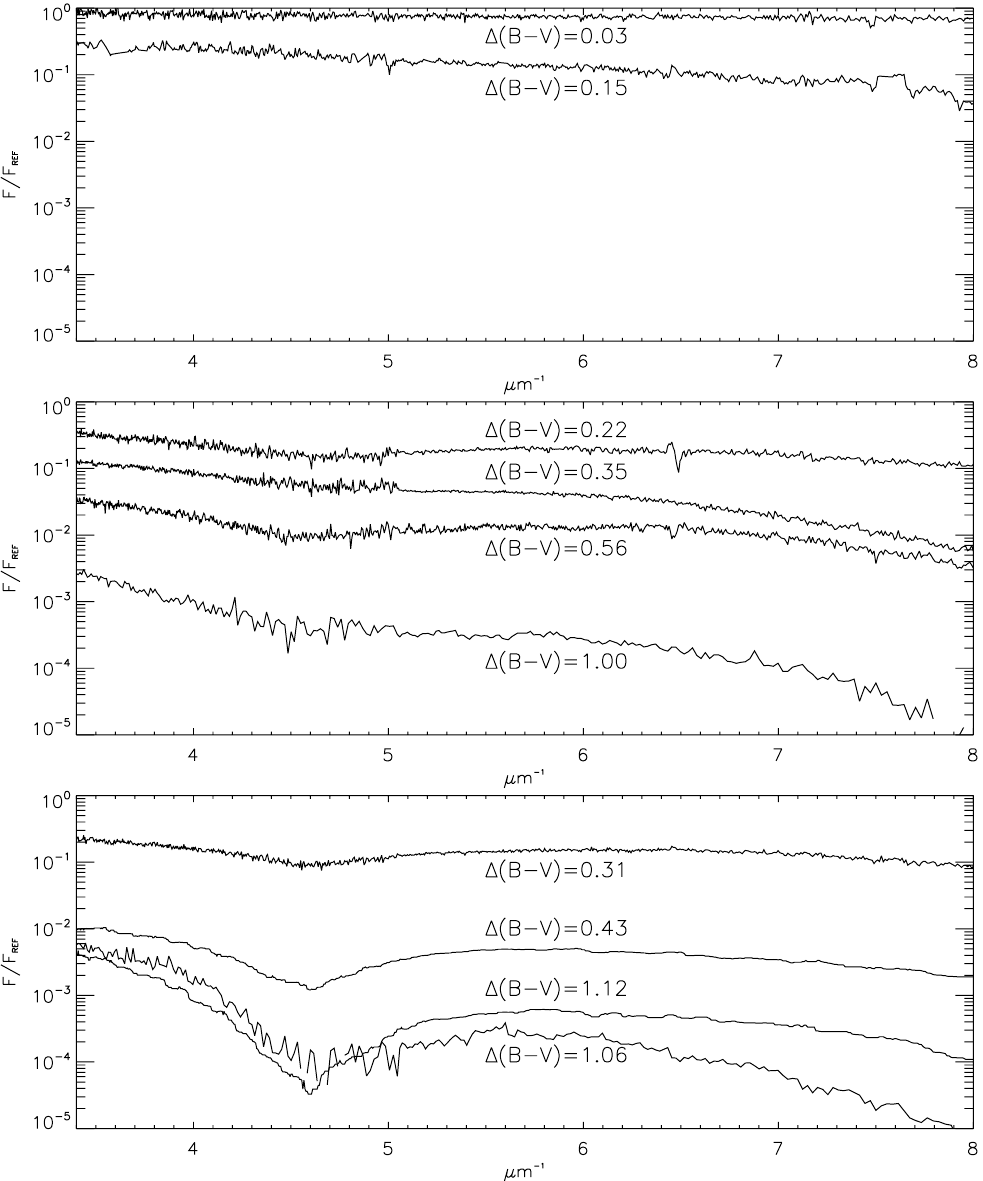}} 
\caption{The three types of observed interstellar extinction curves:  traditional bump-like extinction curves (bottom plot), extinction curves that deviate from linearity in the far-ultraviolet spectrum only (after the bump region, middle plot), linear over the whole visible/ultraviolet spectrum extinction curves (top plot). 
Each spectrum is the ratio of the spectrum of a reddened star to the spectrum of a reference star of low  reddening and of the same spectral type.
The spectra were normalized using the reddening difference $\Delta (B-V)$  between reddened and reference stars, and the \cite{turner14} extinction law for interstellar dust.
Any extinction curve, in the Galaxy or in the Magellanic Clouds, falls into one of the three categories. 
It is clear from the figure that (i) $E(B-V)$ alone is not enough to fix the shape of an extinction curve, (ii) extinction curves with a 2200~\AA\ bump and linear extinction curves anticorrelate.} 
\label{fig:fig1}
\end{figure*}

In the ultraviolet most extinction curves exhibit a singular broad absorption-like band centered close to 217.5~nm, the  2200~\AA\ bump (bottom plot of Fig.~\ref{fig:fig1}).
Curves with a bump have been most studied but, as Fig.~\ref{fig:fig1} clearly demonstrates, extinction curves are not all bump-like.
Some extinction curves do not have a bump, and when this is so the curves are always linear down to the bump region (middle plot of Fig.~\ref{fig:fig1}).
In some cases the curves continue to be linear over the entire  ultraviolet spectrum (top plot of Fig.~\ref{fig:fig1}).

Observation thus separates three classes of extinction curves: bump-like; linear down to the near-ultraviolet (the bump region); and linear over the whole visible-ultraviolet wavelength range (Fig.~\ref{fig:fig1}).
The International Ultra-violet European (IUE) telescope has observed hundreds of reddened stars: with no exception, all extinction curves, whether they are observed in the Galaxy or in the Magellanic Clouds \citep{mc},  fall into one of these three categories.

Extinction curves that are linear in the ultraviolet deserve special attention.
They are more frequently observed in the Magellanic Clouds but are also present in the Galaxy. 
Other than that there is no fundamental difference between Magellanic and Galactic extinction curves \citep{mc}.

In the Galaxy linear extinction curves are observed in two circumstances only: when the reddening is very low and when the obscuring interstellar matter is close to the star (for instance planetary nebulae) (\cite{sitko81,seab81, z05,an13}).
At extremely low reddening (e.g. less than 0.08~mag.) extinction curves are always linear over the whole visible-ultraviolet wavelength range \citep{lc}.

It is clear from Fig.~\ref{fig:fig1} that extinction curves with a bump and linear extinction curves  anticorrelate: the bump disappears when and only when the linearity of the extinction curve in the visible spectrum extends to the near-ultraviolet wavelengths or farther.  
Fig.~\ref{fig:fig1} also shows that reddening is not enough to fix the threshold above which an extinction curve switches to non-linearity: the shape of an extinction curve is not determined by $E(B-V)$ alone.
One or more  additional parameters are needed to fully describe an extinction curve.
\section{Interstellar extinction curves in the standard framework} \label{fitz}
Interstellar extinction curves have been analyzed under the assumption that the light we receive from the direction of a reddened star is the direct  light from the star, extinguished by interstellar particles along the line of sight.
  
Under this assumption three types of particles, each having its own extinction law, are both necessary and sufficient  to account for the variations of the extinction curve with the direction of observation \citep{greenberg00}.
The linear extinction law in the visible part of the spectrum was justified early on by a size distribution of interstellar grains \citep{greenstein38}.
The bump and the far-ultraviolet region are supposed to result from the extinction of light caused by two different species of molecules that still await a formal identification.

\cite{greenberg83} found that the far-ultraviolet rise of extinction curves does not correlate with either the visual extinction or the 2200~\AA\ bump.
Several articles (for instance \cite{meyer81, massa86}) have shown that the bump is only partially, and not univocally, related to the visual extinction.

No connection to the environment has ever been established that would  explain the variations of the extinction in the three spectral domains.
Linear extinction laws exist in nature (atmospheric aerosols), but no laboratory experiment has yet reproduced bump-like extinction curves.

The random variation in the relative abundance of the three types of particles required by standard extinction models was further supported by the mathematical analysis of \cite{fitz88}: if three separate extinction laws (one for the visual extinction, one for the bump, and one for the far-ultraviolet rise) rule interstellar extinction,  interstellar extinction curves must depend on a minimum of six independent parameters in addition to $E(B-V)$.
\section{The CCM paper and its one parameter fit} \label{ccm}
In contrast to the ideas that prevailed earlier (Sect.~\ref{fitz}) \cite{ccm89} (CCM hereafter) reached the remarkable conclusion that the variations observed in normalized extinction curves  correlate over the whole spectrum.
Another study, based on a larger sample of stars, reached the same conclusion a few years later \citep{bondar06}.
CCM deduced  that all normalized extinction curves with a bump can be approximated by a common  fit that depends upon a single parameter, assumed to be $R_V$.
The CCM fit is purely mathematical and empirical.

The CCM paper, a breakthrough in the study of interstellar extinction, also gave rise to several difficulties.
The most obvious one is that it remains embedded in the conventional view that  extinction curves depend on different types of particles, which, as seen above, requires six (instead of one) independent parameters in addition to $E(B-V)$.

A second difficulty arising from the CCM paper is the assumption that $R_V$ is the free parameter of interstellar extinction.
However,  a simple derivation from the fit proves that, in the CCM framework,  $R_V$ must be a constant (=1.46, \cite{rv}).
$R_V$ cannot be both variable, as supposed by CCM, and constant, as implied by the fit.
This antinomy reflects the empirical nature of the CCM fit and the impossibility for $R_V$ to be the free parameter of interstellar extinction.
If interstellar extinction depends upon a single parameter (in addition to $E(B-V)$) and if $R_V$ is not related to this parameter,  $R_V$ must stay constant (independent of direction, \cite{rv, rv2}).

The CCM fit is also not always accurate (fig.~4 in CCM), especially in the 2200~\AA\ bump region, because the fit establishes a one-to-one correspondence  between the bump and reddening (no bump means $E(B-V)=0$ and no extinction, and vice versa).
The fit thus samples one category of extinction curves only,  the curves with a bump (bottom plot of Fig.~\ref{fig:fig1}).
It ignores the linear and linear-down-to-the-bump-region extinction curves of Fig.~\ref{fig:fig1}, and is in manifest contradiction to the fact, already known in 1989, that the size of the bump is tied to $E(B-V)$ but does not depend exclusively on it \citep{savage75,meyer81}.

An improved, still one-parameter-dependent fit,  advantageously replaces the CCM parametrization for bump-like and linear-over-the-whole-spectrum (visible to far-ultraviolet) extinction curves \citep{mc}.
The fit shows that most extinction curves can  be separated into a linear and a  bump-like component.
This separation breaks the univocal dependence of the bump on $E(B-V)$ of the CCM fit, does not increase the number of free parameters, and proves that the normalization of extinction curves, as currently done in most studies, is not a straightforward process and is another reason for the inaccuracy of  the CCM fit.
Like the CCM fit, the new fit lacks a physical basis.
However, it matches all observations with much greater precision, and it applies to nearly all extinction curves except for the minority of intermediate ones (middle plot of Fig.~\ref{fig:fig1}).
\section{The incompatibility between the CCM and Fitzpatrick \& Massa perspectives} \label{ccmfitz}
The CCM and the Fitzpatrick \& Massa fits proceed from two very different approaches.
The latter relies on an assumption about interstellar extinction, the former on an empirical analysis of observations.
The conclusions they reach are opposite and incompatible.

Assuming that interstellar extinction is the superposition of three different extinctions, Fitzpatrick \& Massa  conclude that reddening and six independent (unrelated) parameters are necessary to account for the variety of observed extinction curves.
In contrast, CCM and \cite{bondar06}  consider extinction curves as a set of mathematical functions and search for the minimal dependencies of the curves.
The  CCM conclusion is unequivocal:
\begin{quotation}The most important result presented here is that the entire mean extinction law, from the near-IR through the optical and IUE-accessible ultraviolet, can be well represented by a mean relationship which depends upon a single parameter. We have chosen to use $R_V$ for convenience, but it has no particular physical significance. No doubt other well-observed quantities similarly defined could have served in its stead.

... the deviations of the observations from the mean relation are impressively small ... in the real ISM the processes which modify the extinction at one wavelength also seem to modify the entire mean [normalized] extinction law in a regular way.
\end{quotation}
Few sight lines (7\% out of 417  in \cite{valencic04}) escape the CCM relationship, the exceptions being most certainly related to its purely empirical nature (\cite{mc} and Sect.~\ref{ccm}).

A recent textbook  \citep{draine11} attempts to reconcile the opposing views of CCM and Fitzpatrick \& Massa. 
The author suggests that CCM began with a seven parameter fit $F^7_{CCM}$ and ultimately transformed it into a one parameter  $F^1_{CCM}$ fit, once each of the parameters had been expressed as a function of $R_V$. 

The   \cite{draine11} account of the CCM article is confusing and inexact.
There is no mention whatsoever in CCM of a seven (independent) parameter fit of extinction curves.
No expression  of the Fitzpatrick \& Massa parameters as a function of the CCM parameter was ever found by CCM or anyone else.
Such a finding would  imply a dependency among the Fitzpatrick \& Massa parameters that does not exist: the  Fitzpatrick \& Massa parameters have no relation one to another, and no relation to the free parameter of CCM.
Moreover, the $R_V$ parameter, which CCM adopt as the free parameter of interstellar extinction, is usually taken to be constant in the Fitzpatrick \& Massa fit.
These differences only attest to the discrepancies between the Fitzpatrick \& Massa and the CCM fits, and to the conflict between the  assumptions underlying the Fitzpatrick \& Massa thesis and observation, upon which the CCM finding relies.

Attempts to reproduce a large sample of extinction curves with physically meaningful particles, such as the PAHs, require a much larger number of free parameters than anticipated by observation, 226 (enough to sample the whole visible/ultraviolet spectrum) in a recent ApJ paper \citep{mulas13}.
Although the authors concede  that  the number of free parameters must eventually be reduced from 226 to one, it can be doubted whether this goal is attainable.
\section{Scattered starlight in the spectrum of reddened stars} \label{contra}
The previous sections have shown that several logical deductions can be drawn from interstellar extinction observations.
Interpretations of interstellar extinction curves based on the assumption that we observe  the extinction of the direct light from stars will hardly comport with these conclusions.
Standard three-components models for interstellar extinction face the following difficulties:
\begin{itemize}
\item a three-component model is incompatible with the dependency of extinction curves on reddening and on an additional parameter alone.
\item the free parameter (in addition to $E(B-V)$) of interstellar extinction cannot be $R_V$, which must be a constant.
\item the non-linearity of an extinction curve is intimately linked to the presence of the 2200~\AA\ bump and vice versa. In other words, when (and only when) there is no bump, interstellar dust tends to recover a linear extinction law over the whole spectrum.
\item the extinction cross-section of interstellar dust in nebulae does not have the near-ultraviolet break predicted by three-component models:  the extinction law found in nebulae is continuous and linear over the whole spectrum \citep[and next section]{neb}.
\end{itemize}

In addition, the bump tends to disappear (and the extinction curve becomes linear) when there is very low reddening or when the star is close to the obscuring cloud. 

Besides reddening, the only parameter that is seen to impact the shape of an extinction curve is the distance between the star and the interstellar matter obscuring it, a conclusion that does not make sense if extinction is the only process at work.
This conclusion does make sense, however, if -and this is  the sole alternative to standard explanations of interstellar extinction- what is observed is not only the direct light from the reddened stars, but also a non-negligible amount of scattered starlight.
\section{What type of scattering?} \label{ray}
Two types of scattering are commonly found in nature.
One is scattering by particles with dimensions comparable to the wavelength.
This type is observed to vary as $1/\lambda^p$, with $p$ of order 1, and is strongly forward oriented.
Scattering by aerosols is the obvious example.
The other case involves tiny particles, typically atoms or molecules,  that are small compared to the wavelength.
The scattering is then of Rayleigh type.
It varies as $1/\lambda^4$ (a $1/\lambda^6$ term may also need to be considered \citep{dalgarno62}) and is nearly isotropic.
Rayleigh extinction (and scattering) is far less efficient than dust extinction.

IUE ultraviolet observations of nebulae prove that  light scattered by nebulae in near-forward directions behaves as $1/\lambda^p$ ($p\sim1$) \citep{neb}.
The scattering can safely be attributed to interstellar dust grains and suggests that their linear extinction law at visible wavelengths extends in the ultraviolet.
This result is not compatible with standard extinction models, which all suppose a modification of the size distribution of interstellar grains and a flattening of interstellar dust's extinction law at ultraviolet wavelengths.
Nebulae IUE  observations show no trace of Rayleigh scattering. 

However, there is no way to decompose an extinction curve into  the sum of a component of direct starlight ($\propto e^{-a/\lambda^p}$, $p\sim 1$) and a component of light  scattered by dust grains ($\propto 1/\lambda^p e^{-a/\lambda^p}$, $p\sim 1$).
Forward scattering by dust in the ultraviolet cannot explain the contamination by scattered starlight of  the light received from reddened stars.

\begin{figure}[t]
\resizebox{1.\columnwidth }{!}{\includegraphics{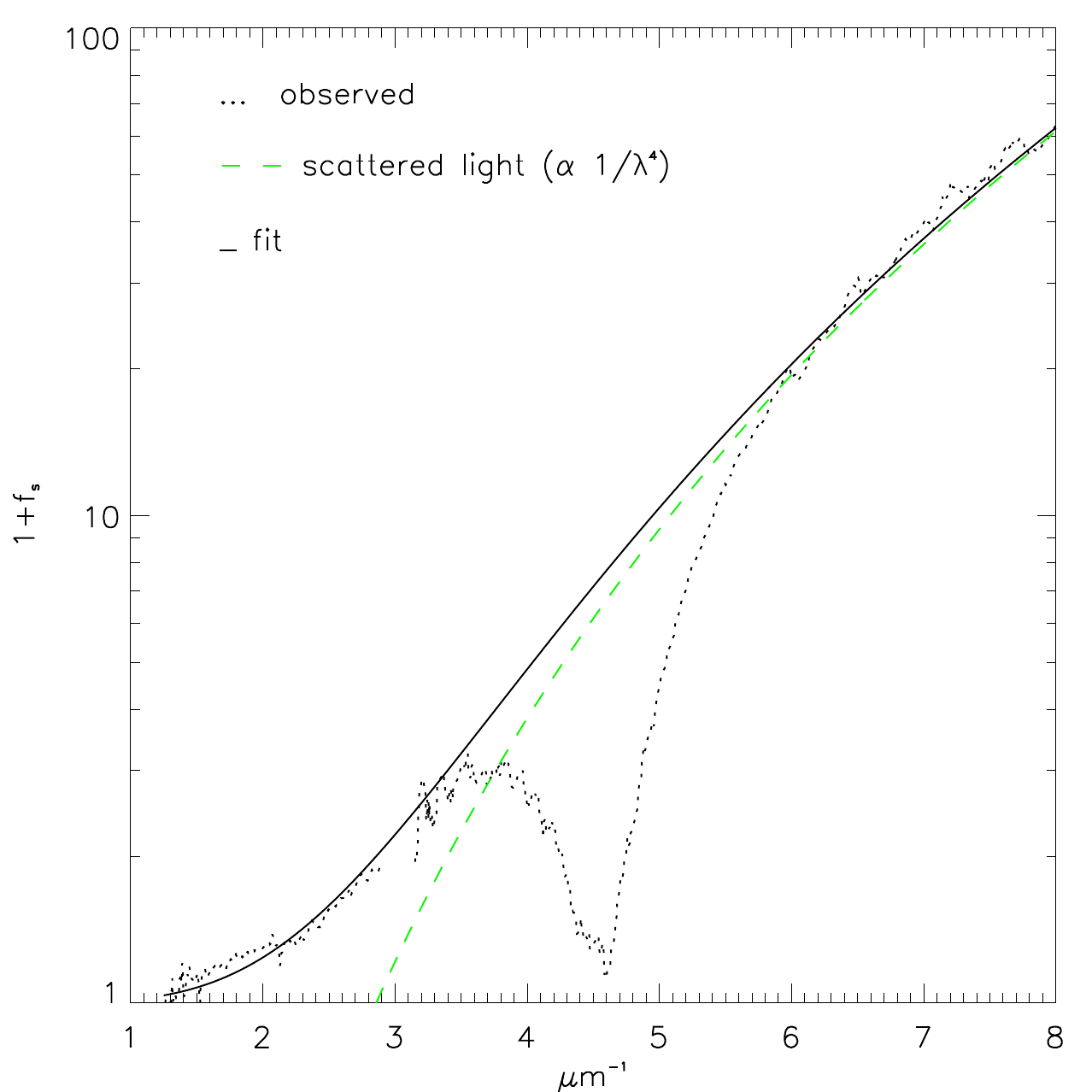}} 
\caption{The ratio $f_s$ of scattered to direct starlight fluxes for HD46223.
The extinction curve of HD46223, corrected for the exponential extinction by dust, is in dots.
The solid black line  ($1+a/\lambda^4$, $a\sim0.015$ if $\lambda$ is expressed in $\mu$m) fits the curve.
The green dashed line represents the continuum of scattered starlight, $f_s=a/\lambda^4$.
} 
\label{fig:fig2}
\end{figure}

The analysis of a complete extinction curve $F/F_0$ of star HD46223, re-constructed from the near-infrared to the far-ultraviolet, shows that the spectral signature of its scattered light component  varies as $1/\lambda^p$,  with $p\sim4$ \citep{hd}.
This finding can be generalized to other directions (Sect.~\ref{ga} and \cite{fultraviolet}).
The scattering is of Rayleigh type, due to gas along the line of sight.
Fig.~\ref{fig:fig2} plots the ratio $f_s$ (dashed line) of scattered to direct light intensities found for HD46223.

There is a problem in observing both dust scattering as close as a few arc-seconds from a star and a significant amount of light scattered by gas in the complete forward direction.
This problem  can be resolved only if the  scattering by gas is coherent in the forward direction.
In this case, the irradiance at the position of the observer should no longer be in proportion to the number of scatterers in the line of sight, $N_X$, but increases as its square, $N_X^2$.
Compared to incoherent scattering, the irradiance is enhanced by a factor $N_H$ (or $N_{H_2}$), that is, by $10^{20}$ to $10^{22}$ (average column densities for HI and H$_2$, as given in table~1 of \cite{friedman11}).
\section{Amount of scattered starlight } \label{isca}
The magnitude of the scattered light component in the ultraviolet spectrum can be estimated by the gap between the observed extinction curve and the continuation of the linear extinction law at visible wavelengths.

The scattered light intensity, expressed in proportion to the direct  light from the star, estimated for HD46223  (Fig.~\ref{fig:fig2}), is of the order of $0.015/\lambda_\mu^4$ ($\lambda$ in $\mu$m)\footnote{This estimation of the proportion of scattered starlight was derived assuming a $1/\lambda$ extinction law for interstellar dust. If the exact law is closer to $1/\lambda^p$, with $p > 1$, the proportion of scattered light should be larger.}.
Both components must be equally extinguished by interstellar dust extinction.

At near-infrared wavelengths, scattered starlight is 2\% of  HD46223's  flux.
The ratio grows to 16\% in the V band and to 30\% in the B band.
In the far-ultraviolet part of the spectrum, scattered starlight overwhelms the star's flux by nearly  two orders of magnitude.
The scattering component can thus be much larger than direct starlight would be without hydrogen in the line of sight.
\section{Consistency} \label{cons}
\subsection{The value of $R_V$ } \label{rv}
$R_V$ is a measure of the relative strength of the extinction by large grains in the B and V bands ($\tau_B/\tau_V=1+1/R_V$), and an indicator of the size distribution of interstellar dust (see \cite{rv2}).
It was recognized as a key parameter for interstellar dust in the 1960s (see \cite{rv}).
Whether it is a constant or varies has been a long-standing debate among observers.
In the wake of the CCM paper most studies have considered  $R_V$ to be variable and the free parameter of interstellar extinction.
Users of the  Fitzpatrick \& Massa parametrization, however, have usually adopted  a constant  $R_V\, =3.1$   \citep{schlafly10}.

The exact ratio of absolute to selective extinction by interstellar grains  is unlikely to be the free parameter of interstellar extinction (Sect.~\ref{ccm}) and is likely to be a constant, $R_V^0$.
Observed  $R_V$ values may differ from $R_V^0$ because of the presence of scattered starlight, which leads $A_V$ to  be underestimated, and $A_B$ to be even more so (the scattering is stronger in the B-band). 
The exponent $p\sim 1$ for the extinction of  direct starlight is smaller than  that for scattered starlight ($p\sim 4$), which implies that observations underestimate the exact $A_B/A_V$ ratios.
Since $R_V=1/(A_B/A_V-1)$, observed $R_V$-values should overestimate $R_V^0$.
The more scattered starlight there is in the spectrum of a reddened star, the greater the overestimation of $R_V$ should be.
These consequences were illustrated  in  \cite{hd} case study of HD46223's extinction spectrum.

 \cite{turner14} did find a minimum, $R_V^0= 2.82$, in observed values of $R_V$. 
The corresponding extinction law $\tau_\lambda\propto1/\lambda^{1.37}$ ($F/F_{ref}\propto e^{-1.14E(B-V)/\lambda^{1.37}}$) agrees with near-infrared observations (fig.~2 in \cite{rv2}).
Incidentally, it also coincides with atmospheric  aerosol observations.
\subsection{Agreement with observation} \label{ga}
The true extinction law of interstellar matter should be well approximated by a quasi-linear law over the whole near-infrared to ultraviolet wavelength range\footnote{At this stage of research, the possibility that a slight variation of the exponent $p$ occurs from one end of the spectrum to the other cannot be excluded.}.
But, on top of the direct light from a reddened star, a telescope intercepts a  component of scattered starlight (by HI or H$_2$ along the line of sight), which explains the departure from linearity of observed extinction curves.
Both direct and scattered starlight are extinguished by interstellar dust and in the same proportion.

Scattered starlight diminishes the impression of extinction and explains why the  departure from linearity that can occur in the visible wavelength range \citep{nandy64} or in  the far-ultraviolet (middle plot of Fig.~\ref{fig:fig1}) is always perceived as less extinction.
When extinction is very low, scattering is very low, and the linear extinction law of interstellar dust emerges clearly over the whole spectrum.
When extinction increases, scattering increases as well, and divergence from linearity is observed.
This effect is especially pronounced in the ultraviolet spectrum.
If the star is close to the scattering material, coherence is diminished or lost (Sect.~\ref{qual}): the linearity at visible wavelengths of the extinction curve tends to be continued in the ultraviolet domain, which is also what is observed.

The shape of an observed interstellar extinction curve is fixed by two parameters: the reddening $E(B-V)$ and a free parameter which regulates the proportion of direct and scattered starlight (thus the size of the 2200~\AA\ bump, Sect.~\ref{pec}).
This free parameter is related to distances between the star, the interstellar cloud, and the observer in a way that remains to be determined.

Extinction in the Magellanic Clouds or in other galaxies need not  be considered as singular.
Linear extinction curves exist in the Galaxy.
Lines of sight towards the Magellanic Clouds sample both  local (Magellanic) clouds, which are infinitely close to the stars from the standpoint of the observer, and Galactic cirrus, relative to which Magellanic stars are infinitely far away.
The coincidence of these two extreme situations along the same lines of sight is the only singularity of  Magellanic extinction curves.
This coincidence, together with the relatively low column densities of the Galactic cirrus in the line of sight,  must  explain why the curves tend to be linear at higher reddening than along standard Galactic sight-lines, and accordingly why they have a reduced 2200~\AA\ bump and smaller DIBs (Sect.~\ref{dib}).

The $1/\lambda^4$ dependence  of the scattered starlight component  of extinction curves can be verified over a large sample of stars \citep{fultraviolet}.
After multiplication by $\lambda^4$, the far-ultraviolet spectrum of an extinction curve with a bump will generally be found to be an exponential with an exponent close to the one deduced from the visual extinction. 
The exact amount of dust extinction at visible wavelengths can thus be retrieved from the far-ultraviolet spectrum of a reddened star, and the relationship between visual and far-ultraviolet extinctions, empirically highlighted by CCM and \cite{bondar06},  finds its physical meaning.
Reddening $E(B-V)$ obtained from the far-ultraviolet spectrum should be larger than the value observed in the visible wavelength range, the latter being slightly diminished because of the contribution from scattered starlight. 
\section{Observation and theory}\label{oth}
\subsection{Van de Hulst formula}\label{vdh}
In Section~4.3 of his book van de Hulst considers the scattering of an incoming plane-wave by a slab of  identical, spherically symmetric particles.
He sums, for the forward direction,  the contributions of the scattered waves to the disturbance at the position of a far-away observer.
He assumes that the particles are separated by distances larger than the wavelength ($n_X\lambda^3\ll 1$, with $n_X$ the density of the particles).

Van de Hulst's formula  leads to a ratio between the scattered and the un-reddened\footnote{ Without the slab of particles in the line of sight.} irradiances  at the observer position  \citep{oc12}:
\begin{equation}
\frac{I_{s}}{I_0}=\left( \frac{4\pi^2 \alpha_X N_X}{\lambda}\right)^2=\frac{3}{8\pi} N_X^2\lambda^2\sigma_X    \label{eq:r1}
\end{equation}
$N_X$ is the column density of the particles, $\alpha_X\sim \frac{9}{2}a^3$ their polarizability, $a$ their mean size.
Cross section $\sigma_X$  is related to the polarizability by $\sigma_X=\frac{8}{3}\pi k^4 \alpha_X^2$  \citep[sect.~6.11]{vdh}.
For molecular hydrogen the cross section comprises terms in $k^6$ and $k^8$ which have, in the ultraviolet spectrum, the same weight as the $k^4$ term \citep{dalgarno62}. 
Eq.~\ref{eq:r1} anticipates a $1/\lambda^2$ dependence for the irradiance of the scattered light if atomic hydrogen is the scatterer, with possible higher powers of $1/\lambda^2$, if it is H$_2$.

For directions with $E(B-V)$ between 0.1 and 1, which are the directions where a bump at 2200~\AA\ is most commonly observed, $N_H$ is between $10^{21}$ and $5\,10^{22}$~cm$^{-2}$ and $N_{H_2}$ typically an order of magnitude less (Table~1 in \cite{friedman11}). 
Condition $n_X\lambda^3\ll 1$ is doubtless verified, since densities should range from a few dozen particles (HI clouds) to a few hundred particles per cm$^3$ (molecular clouds)  \citep{draine11}. 
For $\lambda=1500\,\rm{\AA}=1.5\,10^{-5}$~cm, $\sigma_H=1.2\,10^{-25}$~cm$^2$ and $\sigma_{H_2}=3.4\,10^{-25}$~cm$^2$, $I_s/I_0$ anticipated by Eq.~\ref{eq:r1} is in the range $10^5- 10^6$.
\subsection{ Theoretical irradiance against measured power}\label{alt}
A plane of oscillators creates a disturbance at an observer position that  is fixed by the oscillator strength per unit area (Apps.~\ref{fz} and \ref{fzc}).
A region  scaling with the  first Fresnel zone contributes most to the disturbance.

The oscillator strength per unit area of the interstellar particles responsible for the scattered component of extinction curves is $(2\pi/\lambda)^2\alpha_X N_X$ \citep{vdh,oc12}.
It is $1/\lambda$ for a Huygens-Fresnel wavefront (App.~\ref{fzc}).
The ratio of the two quantities  should give the disturbance of the interstellar dipoles  relative to Fresnel's hypothetic molecules, which is what van de Hulst's formula,  Eq.~\ref{eq:r1} for the irradiance, expresses. 

The formula  contains no information on distances.
This is predictable, since scattering from a first Fresnel zone is independent of the position of the zone between the source of light and the observer (Sect.~\ref{fzc}).

A refined representation of the interstellar cloud, which would also satisfy van de Hulst's (unjustified in the textbook) requirement of a distance larger than the wavelength between the particles,  would be that of a  succession of wavefronts (instead of a single one) along the line of sight, all excited by the same plane wave.
Each wavefront removes a very small fraction of the source energy, and the scattered waves from all wavefronts arrive with the same phase at the observer position.
The expression of the irradiance (Eq.~\ref{eq:r1}), however, should not be modified\footnote{With $N$ identical wavefronts, the surface density of each wavefront is $w=N_X/N$ ($w\lambda^2\leq 1$) and the irradiance due to a single wavefront 1/4 the irradiance due to the first Fresnel zone, $(wS_\lambda)^2 \alpha_X^2 I_0$.
The irradiance for the whole set of wavefronts in the forward direction is $N^2(wS_\lambda)^2 \alpha_X^2 I_0/4=(N_XS_\lambda)^2 \alpha_X^2 I_0/4$, as in Eq.~\ref{eq:r1}.
As long as the thickness $e$ of the cloud remains small compared to the star-cloud and observer-cloud distances, the phase difference at distance $q$ from the star-observer axis does not vary much between the front and back edges of the cloud: there should be no major difference considering the cloud as a succession of Fresnel zones or as a single one.}.
A further simplification of interest would be to limit the cloud to the first Fresnel zones of the wavefronts (for a star far away, a cylinder of radius $\sqrt{\lambda L}$ and depth $e=N_X/n_X$ seen edge on).

These representations underline a similarity with Fresnel zone plates (App.~\ref{fzp}), with   the contributing areas aligned along the line of sight instead of being in an orthogonal plane.
Zone plate theory shows that the contribution of  coherent sources at a focus may greatly enhance the irradiance (as the square of their number) without prejudicing energy conservation.
For the far-field, the increase of the irradiance is balanced by the focalization of the scattered wave along the direction of the source.
The power measured at the focus must remain a limited proportion of the power available for coherent scattering.

\subsection{ Qualitative discussion of the interstellar extinction problem}\label{qual}
For a star infinitely far away, the region of the interstellar cloud that should contribute most to the scattered starlight at the observer position corresponds to an area of order $\lambda L$.
This area is subtended by a very small angle  ($\sim \sqrt{\lambda/L}$), by far much smaller than the resolution a telescope can reach.
Varying the size of the telescope will not modify the proportion of scattered starlight.

As for zone plates, the power measured by the telescope should also be found to be in proportion to the light extinguished within the area of coherence and increase as $N_X$ (instead of $N_X^2$), thus in proportion to $E(B-V)$ and to the size of the Fresnel zone ($\pi\lambda L$) as seen by the observer.
The variations of the average  ratio of the size of the 2200~\AA\ bump to $E(B-V)$, which differs from one Galactic region to another  \citep{meyer81, massa86}, must result from the different distances of these regions from the sun (and not from variations of  interstellar dust composition in different Galactic regions). 

This conclusion is fully supported by the very recent  investigation of  \cite{schlafly16}.
The authors find that the mean observed Galactic $R_V$ value depends on, and increases with, the distance of interstellar clouds (and no other parameter), and they assume that the distance is to be taken from the Galactic center. 
From the standpoint of the present study, however, the distance  is more likely to be from the observer.
The farther away the cloud, the larger its first Fresnel zone will be (with respect to the observer and  for stars supposed to be far away on average), the more scattered starlight there is, implying larger observed mean $R_V$ values (Sect.~\ref{rv}).
These should also correlate with larger 2200~\AA\ bumps.

Deviations from  the average  ratio of the size of the 2200~\AA\ bump to $E(B-V)$ within a Galactic region, on the other hand, are more likely due to variations of  star distances, or to an additional extinction (e.g. circumstellar extinction) in the line of sight. 
When the size of the first Fresnel zone shrinks due to the proximity of  the star, the power available for coherent scattering, the proportion of scattered light in the spectrum of the reddened star, and the 2200~\AA\ bump diminish.

A star may be comparatively close to the interstellar cloud and yet yield large Fresnel zones, meaning that distance star-cloud $D$ is large but that the cloud is much closer to the star than to observer ($D \ll L$), as occurs for Magellanic stars extinguished by local (Magellanic) clouds.
The area of the first Fresnel zone, now of order $\lambda D$, is then small compared to $\lambda L$ and is fixed by the proximity of the star instead of by the observer-cloud distance.

Magellanic stars also have low column density Galactic cirrus (for which the stars are infinitely far away) in the line of sight.
That DIBs are observed at the Magellanic Clouds  redshift (Sect.~\ref{dib}) indicates that the 2200~\AA\ bump and the scattered starlight component of Magellanic extinction curves  should be attributed to Magellanic interstellar clouds rather than to Galactic cirrus.
It follows that the reason why bumps and DIBs are weaker in observations of Magellanic stars is that Galactic cirrus in the line of sight are too light to give appreciable scattered starlight.

The observations discussed in this section seem to imply a symmetry in the roles played by the observer and the star in  determining the amount of scattered starlight, and suggest that the free parameter of interstellar extinction curves is the size of the first Fresnel zone.
Quantitatively, however, there is no straightforward way that I know of to move from the irradiance given by Eq.~\ref{eq:r1} to the power of the scattered starlight that a telescope will measure.
The parallel drawn in this paper between Fresnel and interstellar oscillators, or between the forward scattering by identical dipoles and wave theory, can probably be pushed further.
Abnormally large Fresnel zones, as found on astronomical scales, raise a problem for wave theory, the solution of which should help the analytical treatment of interstellar extinction curves.
\section{Connection between the  diffuse interstellar bands and the 2200~\AA\ bump} \label{dib}
The  fact that the 2200~\AA\ bump and the DIBs concern different wavelength domains and that the bump is a far more prominent feature has fostered a tendency to treat them independently.
For instance, \cite{herbig95}  does not mention the bump in his  review of research on  DIBs.

Studies with a focus on the relationship between the bump and the DIBs have not  always agreed on their correlation \citep{benvenuti89,nandy75, wu81}. 
However, it can be inferred that although there is no strict proportionality between the DIBs and the bump, no proportionality among the DIBs exists either \citep{krelowski87, herbig95, xiang12}.
And, as   \cite{friedman11}  point out, when the DIBs are observed, the bump is observed, and vice versa.
When one is substantially weakened, the other is weakened as well.
These correlations hold for the Galaxy as well as  for the Magellanic Clouds.

\cite{welty06, welty14} find that in the Magellanic Clouds,  DIBs and  the 2200~\AA\ bump are both typically weaker for  LMC and  SMC directions than for Galactic sight lines with similar HI column density, by factors of 9 and 20, respectively. 
\cite{cox07} note that the presence or absence of DIBs in the Magellanic Clouds  is linked to the presence or absence of a 2200~\AA\ bump.
 
In  the Galaxy, there are a few well known sight-lines with anomalously weak DIBs: in  Orion, stars HD36822 ($E(B-V)\sim0.14$), HD37020 (0.37),  HD37021 (0.54),  HD37022 (0.34), HD37042 (0.42), HD37061 (0.52); HD147933
 ($\rho$-Oph, $0.45$); HD24534 ($\chi$~Per, $0.56$); HD29647 ($1$); HD34078 (AE~Aur, $0.52$); HD57061 ($\tau$~CMa, $E(B-V)\sim 0.16$); HD62542 ($ 0.35$); HD148184 ($ 0.5$);  HD223385 (6~Cas,  $0.67$); Herschel~36 ($0.87$) \citep{welty14, dahlstrom13,friedman11, moutou99, snow02}.

In all of these directions the 2200~\AA\ bump is significantly weaker than expected from the reddening.
The ultraviolet extinction curves of Herschel~36 in NGC~6530 and HD29647 in Taurus are remarkable because of the high $E(B-V)$ in these directions and a nearly absent 2200~\AA\ bump \citep{snow80,dahlstrom13}.
The curves are of the intermediate type (middle plot) of Fig.~\ref{fig:fig1}; the weakness of the bump in these curves coincides with the extension of  the linear extinction law at visible wavelength down to the bump region.

A relationship between the DIBs and the distance from the star to the reddening material has also been found for Galactic stars.
\cite{morgan44} associates a  "systematically weakened"  DIB $\lambda4430$ to nebular material in the vicinity of stars.
The tendency for DIB~$\lambda4430$ to increase with star distance also appears on a plot (fig.~6 in \cite{duke51}) of the DIB's strength versus distance moduli for a sample of stars with similar reddening.
From a survey of bibliographical sources \cite{snow95a} conclude  that 
\begin{quote}The general thrust of this literature is that the bands [DIBs] are either weak or absent in circumstellar regions of all descriptions, including both hot star and cool star envelopes.\end{quote}

The Galactic stars with weak DIBs mentioned above are generally bright in the far-infrared spectrum, which is an additional proof that part of the reddening material along the line of sight is close to the star. 
AE~Aur, $\rho$-Oph, and CMa, are known to be close to or embedded in nebulae \citep{vos11,berg66}.
The study by \cite{dahlstrom13} on Herschel~36 separates a foreground  ($E(B-V)\sim 0.38$) and a local  ($E(B-V)\sim 0.49$) extinction with comparable weights in the total reddening of the star.

The diffuse interstellar bands and the 2200~\AA\ bump are concomitant in the spectrum of reddened stars and share the same properties: they tend to increase with the amount of interstellar material, do not depend on the reddening alone, and tend to disappear when interstellar matter is close to the star.
At very low reddening (typically less than  0.08~mag.) neither the bump nor the DIBs are observed.
\section{H$_2$ absorption lines and the Diffuse Interstellar Bands} \label{abs}
 Draine recently commented "It is embarrassing to have to admit that not a single one [DIB] has yet been identified" \citep{draine09}.
In fact, it was remarked as early as the 1970s that many DIBs match allowed transitions from electronically excited states of  H$_2$  \citep{duardo70, duardo71, duardo74, duardo75}. 
The finding was independently made in more recent years and extended by P.~Sorokin and J.~Glownia \citep{sorokin00,sorokin13}.
These theoretical investigations have been supported by laboratory experiments.
These  reveal an absorption spectrum of H$_2$ having "topological resemblance with observed DIB profiles" \citep{hinnen96}, and coincidences between  DIBs and H$_2$ resonances from ground state levels populated at low temperatures \citep{ubachs97, ubachs14}.

Both Duardo and  Sorokin \& Glownia link the H$_2$ transitions that could produce the DIBs to  a two-photon absorption of coherent light: \begin{quote}... a mechanism based entirely upon "passive" two-photon absorption by H$_2$ molecules cannot adequately explain the origin of the DIBs. As will be explained below, we currently assume that some form of coherent light wave structure is present in the H$_2$-containing cloud ... \citep{sorokin98}.\end{quote} 
  \begin{quote}[Two photon absorptions] normally occur with high probability in any medium only if the latter is simultaneously irradiated by intense, ideally monochromatic, light - such as that produced by a laser. \citep{sorokin06}.\end{quote}
Sorokin \& Glownia  were led to the conclusion that the DIBs and the bump are produced in the vicinity of O and B stars\footnote{"... this famous absorption feature [the 2200~\AA\ bump] is \textbf{not} formed in deep interstellar space far away from any stars, as astronomers have generally assumed over the years, but rather originates in the gaseous clouds (containing mostly H$_2$) that often closely surround bright and massive OB stars." \cite{sorokin13}.}, which is untenable with respect of the observed properties of the bump and of the DIBs (previous section and \cite{snow95,snow98}).

That the Sorokin \& Glownia mechanism  cannot be retained should not disqualify their whole model, especially because of the striking correspondence between numerous DIBs and H$_2$ absorption lines.
No other molecule has yet been shown to provide any confirmed DIB identification, and H$_2$ is by far the most abundant molecule in interstellar clouds.

The bump's energy, centered close to 5.7~eV, is about half the energy needed to photo-dissociate  molecular hydrogen \citep[p. 419]{draine11}.
A two-photon absorption, enabled by the coherence of the scattered light along the star-observer direction  as required by the Sorokin \& Glownia hypothesis, could explain the bump, leave unaffected the direct light from the star, and provide the H$_2$ excited state  needed for the absorptions at the DIB wavelengths.

Major difficulties would nevertheless remain in finding the exact selection rules, especially considering that the ground state of hydrogen is gerade, and that according to \cite{ross94} the first gerade-excited state of H$_2$ to be reached in a two-photon absorption is at 12.29~eV.
\section{Further consequences} \label{mis}
\subsection{Star distances} \label{sd}
Scattered starlight in the spectrum of reddened stars affects cosmic distance  estimations because apparent magnitudes are overestimated.
A bias also arises from the fact that standard extinction models tend to underestimate the extinction.

On the one hand, the presence of scattered starlight means that a fair amount of light is introduced into the beam of observation from directions other than the direction of the reddened star.
The star appears brighter than it truly is; its observed apparent V-magnitude $m_V$ is less than what it would be ($m_V^0$) without the scattered starlight.
Neglecting the distinction between $m_V$ and $m_V^0$, the star's distance moduli is underestimated, which implies an underestimation of the star's distance.

On the other hand, standard extinction models assume a smaller extinction than there really is.
Visual reddening $A_V$ and color index $E(B-V)$ are underestimated, while the absolute-to-relative-extinction parameter $R_V$ is overestimated (Sect.~\ref{rv}).
As shown by  \cite{russell14}, even a moderate underestimation of the extinction  leads to photometric distances that can be considerably overestimated. 

A component of scattered starlight in the spectrum of reddened stars, or a deficiency of extinction if the scattered starlight is ignored, both describe the same observation -that is, the gap between observed extinction curves and the extension of the linear visible extinction towards the ultraviolet.
Therefore, if $A_V^0$ is the exact visual extinction, and with the notations defined above: $A_V-A_V^0=m_V-m_V^0$.   

The exact distance moduli for a reddened star of absolute V magnitude $M_V$ is
\begin{equation}
m_V^0-M_V=5\log (d/10\rm{pc})+A_V^0  \label{eq:dis0},
\end{equation}
which can be rewritten as
\begin{eqnarray}
m_V-M_V&=&5\log (d/10\rm{pc})+A_V^0+(m_V-m_V^0)  \nonumber \\
&=&5\log (d/10\rm{pc})+A_V  \label{eq:dis}
\end{eqnarray}
Eq.~\ref{eq:dis} shows that the relative error $\Delta d/d$ is of order $E(B-V)\Delta R_V/5$.

Eqs.~\ref{eq:dis0} and \ref{eq:dis} are different expressions of the distance moduli but lead to  the same value for distance $d$, provided that $A_V$ ($\ne A_V^0$) is correctly estimated.
If standard extinction models do not introduce any specific error on distances, the more scattered starlight there is, the more models need to increase the value of $R_V$ ($\ne R_V^0$, which is a constant).

Empirical methods for the determination of $A_V$  \citep{balazs04, ccm89} and $R_V$ \citep{schlafly16} have been used, resulting in impressively large error margins for photometric distances.
\cite{turner12} for instance, re-considering previous values of $R_V$, finds distances that are cut down by a factor of two.
It can be surmised that, since  $R_V \sim 3$ is generally assumed to be canonical, photometric distances are more likely to be overestimated.
\subsection{The PAH hypothesis} \label{pah}
Poly-Aromatic-Hydrogenated molecules (PAHs) are  part of nearly all three component models for interstellar extinction.
Introduced in the 1980s to explain  the infrared Unidentified Bands (UIBs, \cite{leger84}), PAHs were quickly proposed to explain many of  the interstellar extinction features  over the whole spectrum that were not understood.

PAHs have been supposed to be the carriers of the far-ultraviolet rise of  extinction curves \citep{desert90}, of the 2200~\AA\ bump \citep{steglich10, draine09}, and of the DIBs \citep{duley06}.
They have also been thought to be the source of the Red Rectangle spectral features \citep{witt09}, and, according to \citet{gordon98}, are responsible for some  50\% of the Galactic red light.

PAHs probably have little to do with either the continuum of ultraviolet extinction or the 2200~\AA\ bump (this paper).
Their contribution to the DIBs is disputable \citep{sorokin98, ubachs14,tielens08}.
They certainly do not contribute to the red light of Galactic cirrus \citep{juvela08, pol}, or \emph{a fortiori} to the diffuse Galactic light \citep{z06}. 
Furthermore, should the PAHs exist with the required properties,  the observed fractional abundance of neutral PAHs would be $\sim 10^{-9}$ less than hydrogen \citep{gredel11}, which is extremely low, certainly much lower than what models anticipate (of the order of 10$^{-7}$  \citep{tielens08}).

Despite considerable theoretical and laboratory efforts to match their chemistry to interstellar issues, the very existence of PAHs can legitimately be questioned.
As suggested by the title of the 2010 Toulouse symposium "PAHs and the Universe: A Symposium to Celebrate the 25th Anniversary of the PAH Hypothesis" PAHs remain hypothetical constituents of the interstellar medium.
\section{Conclusion} \label{con}
What does an observer sees when s/he observes a star behind an interstellar cloud?
 If the telescope gathers  only the direct light from the star extinguished by interstellar particles, a minimum of three different types of particles, in proportions that vary from one interstellar cloud to another, each type extinguishing light in a particular spectral range, are necessary to explain the observed visible-ultraviolet interstellar extinction spectrum.
There are logical incompatibilities between this  assumption and observation.

The sole alternative is that scattered starlight is mixed with the direct light we receive from reddened stars and substantially modifies their extinction spectrum.
In addition to the spectrum of a reddened star, the observer retrieves the first order diffraction pattern of hydrogen atoms or molecules in the cloud in the line of sight.

Fresnel's theory can be used to overcome most of the objections that  the attribution of  the non-linearity of extinction curves in the ultraviolet to a scattering process can raise.
It provides the elementary ideas that explain the observed  properties of the scattered light component in the spectrum of reddened stars, its magnitude, and the reason why it becomes appreciable only when distances are extremely large. 
It also indicates that, in addition to column density, the key parameter of interstellar extinction is related to the size of the first Fresnel zone at the cloud position, as seen by the observer.
However, diffraction from very large Fresnel zones still creates a challenge for the theory, which does not provide a straightforward fit for interstellar extinction curves.

Over the years this alternative interpretation of the continuum of extinction curves has been confirmed by different types of observation.
These include the ultraviolet spectrum of nebula, which does not indicate a discontinuity between the visible and the ultraviolet spectrum of the scattering cross-section of interstellar dust; the ultraviolet extinction law of stars with very little reddening, which is linear over the whole spectrum; the analysis of the relationship between the extinctions in the visible and the far-ultraviolet parts of the spectrum; and the study of  Magellanic extinction curves, which tends to prove the universality of the extinction law of interstellar matter. 
That observed values of $R_V$  depend on the distance of interstellar clouds is another finding that standard extinction models can hardly accommodate.

Three implications are worth noting:
\begin{enumerate}
\item{
The three major features of interstellar extinction (the deviation of the continuum from linearity in the ultraviolet spectrum, the 2200~\AA\ bump, and the diffuse interstellar bands) are intimately connected. They have the same properties (e.g. they cancel at very low reddening or when the interstellar matter is close to the reddened star) and are concomitant.  
The bump has the peculiarity of being  a selective extinction of the scattered starlight component of extinction curves.
}
\item{
Three-component models and the assumption that scattered starlight contaminates the spectrum of reddened stars are mathematically equivalent, insofar as distance estimations are concerned.
Classical (no scattered starlight) derivations of cosmic distances underestimate the true extinction and need to overestimate the $R_V$ parameter.
The adoption of the canonical $R_V=3.1$ value with a standard extinction model, for instance, can lead to overestimating photometric star distances.
}
\item{This revision of  interstellar extinction challenges the basis of traditional interstellar dust models, i.e. the separation of extinction curves into three distinct wavelength domains, the justification of which has led to the search for  hypothetical molecules such as the PAHs.
The existence  (with the required properties) of these molecules is unnecessary and doubtful.
They are likely to meet the  fate of Fresnel's "mol\'ecules eth\'er\'ees" and other such unsuccessful quests in the history of physics  for particles that do not exist.
This is all the more likely for particles whose density should  be  $10^{-9}$ the density of H$_2$ \citep{gredel11}, in  a medium, the interstellar medium, that is a near perfect vacuum.}
\end{enumerate}

In the framework developed in this paper, I have not questioned the nature of the interstellar dust grains responsible for the extinction of starlight. 
It is more likely to be the  size distribution, rather than  the chemical composition of the interstellar grains or molecules, that produces the direction-independent, quasi-linear extinction law over the whole near-infrared to far-ultraviolet spectrum.
Whether interstellar dust  is a mixture of graphite particles or  silicates \citep{hoyle62,hoyle69}, or other particles, does not affect the present work.

Hydrogen (atomic and/or molecular) should be the major and probably unique agent of the remarkable properties of interstellar extinction that are observed on top of the linear dust extinction: the ultra-violet departure from the linear dust extinction law, the 2200~\AA\ bump, and the DIBs.
Item~1 above implies that DIB research cannot be conducted independently from a more general reflection on the two other major features (the continuum and the bump) of interstellar extinction.
Sect.~\ref{abs} recalled that molecular hydrogen is the only molecule to which a large number of DIBs can be assigned, a fact that would be worth  investigating more deeply.
It was suggested in this paper that the 2200~\AA\ bump was a two-photon absorption by molecular hydrogen of coherent scattered starlight. 
From the excited state, molecular hydrogen absorption at the DIBs frequencies becomes possible.
\section*{acknowledgement}
I thank David Turner for pointing a mistake in a previous version of Sect.~\ref{sd} (on star distances), and Laleh Safari, Filippo Fratini, Kari J\"{a}nk\"{a}l\"{a}, for introducing me to the difficulties the bump and the DIBs can raise in the context of molecular hydrogen theory.

{}
\appendix
\section{Diffraction theory as the basis of coherent forward  Rayleigh scattering}\label{cs}
Fresnel's theory, the first theory to account for diffraction phenomena, lies behind scattering theory. 
The basic idea (Huygens' principle) supposes the existence of a medium of small oscillators  ("mol\'ecules  \'eth\'er\'ees") which transmit the  light vibration, preferentially in the forward direction.
Light propagates through the excitation of the successive layers (wavefronts) of  oscillators centered on the source of light.
The specific nature of these oscillators and the characteristics of the medium were never clarified, for the good reason that they do not exist.

Modern presentations of wave theory rely on the estimation of  the contribution of a small area $\delta S$ of a wavefront to the irradiance at an observer's position, or focus, $P$.
Area $\delta S$ (at distance $D$ from the source and $l$ from the observer) needs to be neither too small (Sect.~\ref{osd}), nor too large (to avoid appreciable phase-lags within $\delta S$) compared to the wavelength.
The contribution $u_{\delta S}$ of $\delta S$ to the disturbance $U_0$ at $P$ is  fixed by the size of the area (the number of oscillators is proportional to the area) and by the path-difference $\Delta_{\delta S}$ between the trajectory  light follows from the source to the observer passing by $\delta S$, $D+l$, and the straight  source-observer distance $D+L$.

Dismissing the $e^{i\omega t}$ time-dependence, the disturbance $u_{\delta S}$ consists in an amplitude and a phase-lag term.
The amplitude, $\epsilon_\lambda A\delta S/(lD)$, is proportionate to the radiation field at the wavefront location ($A\delta S/D$) and propagates as the inverse-distance $1/l$ from $\delta S$ to the observer.
The proportionality coefficient $\epsilon_\lambda$ depends on the wavelength and stands for the strength per unit area  and unit radiation field of Fresnel's oscillators in the forward direction.
  
The phase-lag term $e^{i\varphi_{\delta S}}$ of $u_{\delta S}$ differentiates the contributions of different areas according to their position on the wavefront.
If $r$ is the distance from $\delta S$ to the source-observer axis
\begin{eqnarray}
\Delta_{\delta S}&=& \frac{r^2}{2} \left(\frac{1}{D}+\frac{1}{L}\right) \label{eq:delta}\\
\varphi_{\delta S}&=&\frac{2\pi}{\lambda}\Delta_{\delta S}=\pi r^2 \frac{D+L}{\lambda DL} \label{eq:ph} \\
u_{\delta S}&=&\frac{A}{D}\frac{\epsilon_\lambda}{L}e^{i\varphi_{\delta S}}\delta S 
 \label{eq:u}
\end{eqnarray}
The irradiance that area $\delta S$ alone would induce at $P$ is in proportion to $\delta S^2$ although the power crossing the area is  proportional to $\delta S$.
Therefore, $N$ identical neighboring areas with negligible phase differences at $P$ contribute as $N^2 \delta S^2$ to the irradiance at $P$, while the power crossing the areas increases as $N \delta S$.
This implies a concentration of power at the focus and a depletion elsewhere that underlie diffraction phenomena.
\subsection{The first Fresnel zone}\label{fz}
\subsubsection{Definition}\label{fzd}
The first Fresnel zone is the disk centered on the direction of the source within which contributions of a wavefront cooperate constructively at the position of the observer.
Light paths between the source and the observer passing through any two points within the zone differ by less than half the wavelength ($\varphi_{\delta S} < \pi$, $\Delta_{\delta S}<\lambda/2$).
Successive Fresnel zones are defined by rings of constructive interferences (from positions within the ring) at the observer position.

At distance $L$ from the observer and $D$ from the source, the radius and area  of the first Fresnel zone are
\begin{eqnarray}
r_\lambda&=&   \left({\lambda \frac{DL}{D+L}} \right)^{0.5}    \label{eq:rl}\\
S_\lambda&=&    \pi\lambda \frac{DL}{D+L}  \label{eq:sl}
\end{eqnarray}
For a given distance between the source and the observer, $S_\lambda$ is small close to the source and close to the observer, and maximal   ($S_\lambda=\pi\lambda L/2$) in-between.
But the largest size a Fresnel zone can reach at a given distance from the focus, $S_\lambda=\pi\lambda L$, is obtained when the source of light is a plane wave (the source  is infinitely far away and the wavefront is a plane).

The order of magnitude for the radius of a first Fresnel zone  is typically a few millimeters in laboratory experiments, 100~m within the solar system, 1000~km a few hundred parsecs away.

The Fresnel zone of order $j$ is  the region of the wavefront within which the phase shift $\varphi_{\delta S}$ is in-between $(j-1)\pi /2$ and $j\pi/2$.
It is limited by the rings of radius $r_j$ and $r_{j+1}$ with
\begin{eqnarray}
r_1&=&r_\lambda   \nonumber\\
r_j&=& j^{0.5}r_\lambda     \label{eq:rj} 
\end{eqnarray}
All Fresnel zones have the same area ($\pi r_\lambda^2$), but the thickness of the ring they delimit decreases quickly (as $r_\lambda/\sqrt{j}$) with the increasing order  $j$ of the zone.

\subsubsection{Contribution of a first Fresnel zone to the irradiance}\label{fzc}
Should a first Fresnel zone be divided into $N$ rings of equal area $\delta S=S_\lambda/N$, the phase-lag of ring $j$ is  $\varphi_{j}=j\pi /N$.
The disturbance at $P$ due to the first zone alone is (Eq.~\ref{eq:u} and $N\gg 1$)
\begin{eqnarray}
U_1&=&\sum_{\tiny{j=0}}^{\tiny{N-1}} u_j = \frac{A\epsilon_\lambda}{DL}\delta S\sum_{\tiny{0}}^{\tiny{N-1}} e^{i\varphi_{j}} \nonumber \\
&=&\frac{2i}{\pi}\frac{A\epsilon_\lambda}{DL}S_\lambda=2i\lambda \frac{A}{D+L}\epsilon_\lambda \label{eq:u10}\\
&=&2i\lambda  \epsilon_\lambda U_0,
 \label{eq:u1}
\end{eqnarray}
since $\sum_{\tiny{0}}^{\tiny{N-1}} e^{i\varphi_{j}}=\frac{2i}{\pi}N$.
Disturbance $U_1$ from a first Fresnel zone is independent of the position of the wavefront between the source and the observer and is, as expected, proportional to disturbance $U_0$ at P (the observer's position). 

Fresnel argued  that a limited part (characterized by the size of the first zone) of the wavefront around the source-observer axis contributes most to the irradiance at $P$, since
a small solid angle issued from the observer position  intercepts an increasingly large number of Fresnel zones  as it shifts away from the axis, and their  contributions tend to cancel out.
In addition, the disturbance  from any half of a Fresnel zone, except for the first half of the first Fresnel zone, is annihilated by the half from one of the two nearest zones. 
Accordingly, disturbance $U_0$ at $P$ should be the disturbance due to half the first zone alone and half the disturbance due to the whole first zone.

Fresnel's argument neglects the phase variations within a zone, and was eventually refined \citep{born, hecht}:
the amplitude of the  disturbance at $P$ is half the amplitude of the disturbance from the first zone and $1/\sqrt{2}$ times the disturbance's amplitude due to half the first zone.
The irradiance at $P$ is 1/4 the irradiance due to the first Fresnel zone and half the irradiance due to half of the zone.

Equations (Eqs.~\ref{eq:delta} to \ref{eq:u1}) and most reasonings thus far  apply to any set of identical dipoles, regardless of the particular case of light diffraction.
The value of $\epsilon_\lambda$ is fixed by the nature of the oscillators.

For Fresnel oscillators $|U_1|=2|U_0|$ implies  $\epsilon_\lambda=1/\lambda$ and a quarter of a period phase shift between $U_1$ and $U_0$ (Eq.~\ref{eq:u1}).
From Eq.~\ref{eq:u},
\begin{eqnarray}
U_1&=&\frac{2Ai}{L+D}=2iU_0  \label{eq:u1f}\\
u_{\delta S}&=&\frac{A}{\lambda LD}e^{i\varphi_{\delta S}}\delta S 
 \label{eq:uf} 
 \end{eqnarray}
Eq.~\ref{eq:uf} gives the disturbance at $P$ per unit area of the wavefront, $u_{\delta S}/\delta S$.
 
If $N$ identical small areas $\delta S$ cover the first Fresnel zone, the disturbance they create at the observer position $P$ will necessarily be $U_1=\Sigma u_{\delta S}=\frac{2i}{\pi}S_\lambda\, |u_{\delta S} |$.
The irradiance at $P$ is 1/4 the irradiance $I_1$ due to the first Fresnel zone alone, and equals the irradiance due to any sub-set  of the first zone with identical areas, as they had no phase lag at $P$, divided by $\pi^2$:
\begin{equation}
 I_0=\frac{1}{4}I_1=\left(\frac{1}{\pi}\right)^2N^2| u_{\delta S}|^2=\left(\frac{1}{\pi}\right)^2 S_\lambda^2 \left|\frac{u_{\delta S}}{\delta S}\right|^2
 \end{equation}

A remarkable property of first Fresnel zones is that their contribution to the disturbance (equally to the irradiance) at $P$ depends neither on the size of the zone (its position  between the source and the observer) nor on the wavelength.
The importance for optics of first Fresnel zones was  acknowledged by  \cite{vdh}  in the following  terms: 
\begin{quote}The amplitude at a distance $l$ beyond a plane wave front is such as if an area $\lambda l$ of the wave front contributes with equal phase and the remaining part of the wave front not at all ... A pencil of light of length \emph{l} can exist only if its width at its base is large compared to $\sqrt{\lambda l}$.\end{quote}
\subsection{Zone plates}\label{fzp}
Irradiance at a focus is amplified by a factor of 4 if the light from the first Fresnel zone alone is selected.
The necessary counterpart of the increase of power at the focus is that it is concentrated on a small disc, the Airy disk, around the focus.
 
 If $I_1(q)$ is the irradiance in the focal plane at distance $q$ from the axis, due to diffraction of a plane wave by the first Fresnel zone ($r_1=\sqrt{\lambda L}$), 
\begin{eqnarray}
I_1(q)= I_1(0) \left[2\frac{J_1(z)}{z}\right]^2
 \label{eq:airy1} \\
 \left(z = kr_1\frac{q}{L}= 2\pi\frac{q}{ \sqrt{\lambda L}}\right) \label{eq:z1}
 \end{eqnarray}
$J_1$ is the Bessel function of the first kind of order 1.

The  first minimum of $|J_1(z)/z|$ is for ($z=3.83$, $q_1=\frac{1.91}{\pi}\sqrt{\lambda L}$) and limits the Airy disk of area $Q_1=\frac{3.65}{\pi}\lambda L$.

The power $P_1$ across the spot is
\begin{eqnarray}
P_1&=&I_1(0) \int_{z=0}^{z=3.83}\! \left[2\frac{J_1(z)}{z}\right]^2\,2\pi q\mathrm{d}q
 \nonumber \\
 &=&\left[ \frac{\lambda L}{\pi } \int_{0}^{3.83}\! \frac{J_1(z)^2}{z}\,\mathrm{d}z\right] \,I_1(0) \nonumber\\
 &\simeq&\frac{0.4}{\pi}  \lambda L\,I_1(0) = \frac{1.6}{\pi}  \lambda L\,I_0
 \label{eq:airyp}
 \end{eqnarray}
In contrast to the irradiance ($I_1=4I_0$ at the focus), $P_1$ depends on the area of  Fresnel zones.
It remains less than the power $\pi\lambda L\,I_0$ crossing the Fresnel zone.

Irradiance at the focus can be increased even more by adding the cooperative disturbances from Fresnel zones of the same parity.
A zone plate comprising $N$ zones of the same parity has a theoretical irradiance at the focus $I_N(0)$ $=N^2I_1(0)$ $(=4N^2I_0)$. 
The irradiance is amplified by $4N^2$.
Since the 1950s  zone plates with  hundreds of Fresnel zones have been fabricated in the laboratory.
A zone plate with 500 zones for instance amplifies the irradiance by a factor $10^6$.

The counterpart to this great increase in the irradiance is a concentration on a limited area around the focus, $s_N$,  of the power crossing the focal plane.
Should it be supposed that the power across $s_N$ is nearly $I_N(0)s_N$ and remains a non-negligible part of the power $P_N= NP_1=NS_1I_0$ flowing through the zone, $s_N$ must be of the order $S_1/N$ and must decrease as $1/N$ (since $I_N(0)=4N^2I_0$).

Zone plate theory  \citep{boivin52,childers69} proves that  the distribution of light in the focal plane of a zone plate continues to be well approximated by the Airy diffraction pattern of a circular aperture having the size of the zone plate\footnote{The property can be understood because  when  $N$ is large Fresnel rings become extremely thin.} (radius $a=r_{2N-1}\simeq \sqrt{2N\lambda L}$, area $S_{2N}=2\pi N\lambda L$).

Accordingly, the irradiance in the focal plane at distance $q$ from the axis is
\begin{equation}
I_N(q)\simeq I_N(0) \left[2\frac{J_1(z)}{z}\right]^2
 \label{eq:airyn}
 \end{equation}
with
\begin{equation}
 z = kr_{2N-1}\frac{q}{L}= 2\pi q\sqrt{\frac{2N}{\lambda L}} \label{eq:zn}
 \end{equation}
The radius of the bright focal spot is obtained for $z=3.83$ and worth $q_N= \frac{1.91}{\pi}\sqrt{\lambda L / (2N)}$. 
Its area, $Q_N=\frac{3.64}{\pi}\lambda L/N$, diminishes as $1/N$ (as anticipated above), and increases with the size of the first Fresnel zone.

The power $P_N$ across the focal spot is calculated as for Eq.~\ref{eq:airyp}
\begin{eqnarray}
P_N &=&\left(\frac{0.54}{\pi^2}Q_N\right)I_N(0) \nonumber \\
&=&\frac{2.15}{\pi^2}  \,N^2 Q_N\,I_0= \frac{0.4}{\pi^2}  \,S_N\,I_0
 \label{eq:pn}
 \end{eqnarray}
$Q_N\,I_0$ is the power that would be measured across the bright spot if there were no zone plate; $S_N\,I_0$ is the power crossing the zone plate.

The power $P_N$ an observer measures over $Q_N$ varies as $N^2$ times the power that would be measured without the zone plate. 
It nevertheless remains in proportion to the available power: the $\propto N^2$ growth of the irradiance at the focus is compensated by the $1/N$ decrease of the area of the bright central spot.
Energy conservation is ensured, since the power crossing $Q_N$ remains  in a constant ratio ($\ll 1$) with the power on the zone plate.

Diffraction by a zone plate tends to align the direction of the diffracted light with the direction of the source wave, and to focalize the available power within a solid angle $\Omega_N=Q_N/L^2=2\lambda/(\pi N L)=2\lambda^2/S_N$.
\subsection{Diffraction by obstacles (particles or apertures) with small dimensions compared to the wavelength} \label{osd}
Lorentz's atomic dipoles may be what matches best the idea of Fresnel's oscillators \citep{stone}.
Unlike areas $\delta S$ with a size comparable to the wavelength, dipoles have a nearly isotropic phase function  (within a factor of 2 between the back and forward directions). The disturbance due to a dipole depends on the wavelength as $1/\lambda^2$ (instead of $1/\lambda$), and the irradiance as $1/\lambda^4$ \citep{vdh,stone}.

Theory also shows that the  extension of Eq.~\ref{eq:uf} fails when applied to apertures with dimensions that are small compared to the wavelength \citep{bethe44}.
A very small  area diffracts light with the same $1/\lambda^2$ wavelength dependency of the disturbance as a dipole does.
The disturbance due to a dipole or a small  aperture is $\propto \lambda^{-2}a^3$ (irradiance  $\propto \lambda^{-4}a^6$), where $a$ is the typical size of the dipole/aperture.

To ensure continuity between the two regimes (small or large apertures/obstacles), several small diffracting apertures/obstacles covering a larger area of order $\lambda^2$ must yield the forward disturbance of Eq.~\ref{eq:uf}.
This can only occur if  a thickness of order $\lambda$ is attributed to Huygens-Fresnel wavefronts, and if the $a^3$ term in the disturbance created by a dipole is interpreted as corresponding to its volume $\delta V$.
The disturbance created at a focus by a set of apertures or obstacles smaller than the wavelength is then $\delta u\propto \lambda^{-2}\delta V$ (since $\delta S\sim \delta V/\lambda$ in Eq.~\ref{eq:uf}). 

If a thickness  $ \lambda$ is attributed to wavefronts, and if Fresnel's oscillators  (radius $a_F$, density $n_F$, polarizability $\alpha_F\simeq \frac{9}{2}a_F^3$, column density of a wavefront $N_F=\lambda n_F$) had any existence at all,  the oscillator strength per unit area of a wavefront would be $(2\pi/\lambda)^2\alpha_F  N_F=(2\pi/\lambda)^2\alpha_F \lambda n_F$ on the one hand (Sect.~\ref{alt}), and $1/\lambda$ on the other. 
Their radius and density would be related by: $18\pi^2n_Fa_F^3\simeq1$.

\end{document}